\documentclass[runningheads]{llncs}

\usepackage[dvipsnames]{xcolor}

\usepackage{soul}
\usepackage{graphicx}

\usepackage{hyperref}
\usepackage{array}
\newcolumntype{P}[1]{>{\centering
\arraybackslash\hspace{0pt}}p{#1}}

\usepackage{rotating}

\usepackage{makecell}

\usepackage{float}
\restylefloat{table}

\usepackage{booktabs}
\usepackage{adjustbox}
\usepackage{enumitem}

\begin{document}
\title{Change Logging and Mining of Change Logs of Business Processes - A Literature Review}

\titlerunning{A Literature Review on Change Mining and Logging in PAIS}
\author{Arash Yadegari Ghahderijani \and Hande Naz Turgay \and  
Dimka Karastoyanova}
\authorrunning{A. Yadegari, H. Naz Turgay and D. Karastoyanova}
\institute{Information Systems Group, University of Groningen, The Netherlands}
\maketitle              % typeset the header of the contribution
\begin{abstract}
    \textit{Context:} Change mining enables organizations to understand the changes that occurred in their business processes. This allows them to enhance their business processes and adapt to dynamic environments. Therefore, change mining is becoming a topic of interest for researchers, scholars, and practitioners. \\
    \textit{Objective:} Motivated by the goal of establishing the state of the art in this area, this paper aims to investigate the literature in change   logging and mining in process-aware information systems, provide an overview of the methods that are used in the existing publications, and identify gaps in the research on the topic of logging and mining process changes. \\
    \textit{Method:} A literature review is conducted with the objective to identify and define methods to mine, store, and record changes in business processes. From 1136 publications, we selected 6 papers related to changes in business process and extended the list to 9 papers by including the relevant articles referenced by the papers that we selected originally. \\
    \textit{Results:} In answer of our research questions, we have identified two classes of change mining methods, two ways of recording the changes into change logs, five formats for change log representation, and four objectives to be learned from changes. \\
    \textit{Conclusion:} The literature review provides a summary of existing change mining and logging methods in process-aware information systems and identifies a number of research gaps in the area.

\keywords{
Process Change Mining \and Change Logs \and Business Processes \and Process-Aware Information Systems \and Business Process Management \and Literature Review
}
\end{abstract}

\section{Introduction}
\label{section:introduction}
Change mining is carried out by organizations to examine the details of past changes in their business processes and provides them with information and insights that helps them to plan and evaluate possible future changes. The visibility of changes in business processes can help organizations to manage and optimize their systems and use of resources bringing in long term benefits. 
The topic of process change has been considered in the BPM field from several perspectives, from logging and recording the changes, through the discovery/mining of changes (also known under the term adaptations)~\cite{purging}, to the mechanisms for enabling change manually, automatically or a mixture of the two~\cite{applyingchange,handlingsudden,enhancing,anewframework}. The information systems enacting business processes can be natively process-aware or be general purpose ones and the type of system influences the solutions towards discovering and managing process change. With the staggering improvements process mining and machine learning have already bring in into the field of BPM, process performance improvement is becoming an even more exciting research area. One open direction of research is the search for best practices to recommend and enact changes with highest positive impact on process performance while the processes are being executed.
In the context of that endeavour, our goal in this paper is to establish the state of the art on how process changes are recorded during process execution and what mining approaches are used for identifying change based on the recorded information about process instance execution. Furthermore, we are interested in providing an overview of the different types of results the change mining approaches provide. 

Based on that, we aim at identifying directions for future work. The methodology we have selected to achieve this goal is systematic literature review (SLR)\cite{Kitchenham2007} as this methodology fits best our objectives.

The contributions of our work are 1) a literature review of the state of the art and 2) identification of gaps in the research on the topic of logging and mining process changes.
Our study will include results related to runtime process changes in both PAIS and general purpose information systems. We include in our search results reporting on if and how process change information is recorded, how changes are identified and logged, what approaches are used to mine the logged changes and in what form are the mined changes represented. We acknowledge that approaches for process drift detection and anomaly detection are related to process change and have been discussed in multiple works from literature. Due to this abundance and aiming at shaping our study to a manageable scope, we considered only works whose target were changes made with the purpose to improve process performance, rather than being anomalies or result of process drift.

The remainder of the paper is organized as follows. In Section \ref{section:methodology}, the research methodology to conduct the literature review is explained in detail, including the research questions we identify. Section \ref{section:results} presents the extracted data and results by synthesizing the information from the literature that we collected to answer the research questions. Section \ref{section:discussion} provides a discussion of the results and our findings. Finally, Section \ref{section:conclusion} concludes the review.

\section{Background Information}

In this section we provide an overview of several distinct but related research areas in the field of business process management (BPM) with the goal to provide useful background information to the reader of this study.
\subsection{Business Processes and Process Aware Information Systems}
\label{section:bp}
Business processes can be defined as a series of activities that address a business objective in an organization and allow business activities to be executed efficiently. Business process models are sets of relationships between business activities, which depict the possible sequences of events\footnote{An event is the execution of a business activity.
} that may occur during execution time. 

Business process management (BPM) is a management approach that combines information technology with management sciences to model and enhance business processes \cite{paisdesign}. \\
\indent The modification of a business process to meet new business requirements is called process change. 
Process changes can occur at  build-time/design time or run-time of a business process and result in the addition of new features through modifications to the process elements. Process changes must be analyzed and dealt with as the validity and performance of the original business process models, as well as their instances, might be affected by them.

A process-aware information system (PAIS) is a special kind of information system that handles the management and execution of processes, which contain people, applications, and data sources based on process models and needs to be able to deal with ambiguity, unexpected situations and changes in  environment \cite{flexibility} \cite{lessonstobe}. PAISs are commonly used for workflow management, case handling, and in enterprise information systems \cite{lessonstobe}. 

\indent Many works have been conducted in order to make PAIS more flexible, which resulted in the emergence of adaptive process management approaches (e.g. ADEPT, CBRFlow, WASA) that allow users to dynamically change process models to address changed requirements. Adaptive PAIS enable the dynamic application of changes to different process aspects, such as control-flow and data-flow, at multiple levels, such as process instance and process type. Particularly, the application of ad-hoc changes at the instance level enables the adaptation of single process instances to dynamic conditions \cite{adeptflex}, \cite{DBLP:conf/otm/WeissSHK14,DBLP:conf/bpm/SonntagK11}. Some adaptive PAIS  record ad-hoc changes in so-called change logs \cite{purging} which, in turn, provides more meaningful log information than the log information that can be retrieved from traditional PAIS. 

One approach that has been developed specifically to allow for change and reuse are the so-called configurable processes \cite{semantic}. A configurable business process model combines variants of a business process in a single model in order to simplify the management of business process variants \cite{gottschalk}. Configurable business process models can be represented by modeling languages such as C-EPC \cite{basisfor} and C-BPMN \cite{extending}, which allow the generation of process variants from a configurable process model. \\
Business process variants can be created from the configurable process model by configuring the variation points. Variation points are the places in which the variation happens in the business process model and each one has multiple alternatives called variants. A new process variant is generated through the assignment of variants to variation points \cite{gottschalk}. There are three steps to manage variability in a configurable business process \cite{semantic}: Modeling,  Configuring and  Evolving.

\indent A business process or a family of business processes can evolve and change to fulfill new requirements \cite{comperative}. These changes are registered as events on event logs which are used as a source for all process mining and change mining \cite{manifesto} methods.

\subsection{Process Mining and Change Mining}
\label{section:pm}
Process mining is the extraction of information from event logs, which are stored in information systems, to explore, track, and enhance business processes. Process mining covers perspectives such as \cite{manifesto} control flow perspective, organisational perspective, case perspective and time perspective focusing on identifying different aspects of processes based on process event logs.

%The input for process mining methods is event logs.
In an event log, the information about how events are ordered should be available and each event should be connected to a process instance and a specific activity. In most event logs, the information about fields such as the performer of the event, the timestamp of the event, or data recorded with the event is present as well \cite{usingprocessmining}. 

\indent Changes occur by “applying a sequence of change primitives or operations to the respective process graph” \cite{purging}. Changes can be applied to Business Processes and include addition, deletion, or modification of an activity, resource, or data of a process \cite{static} or to Configurable Business Processes, where the  addition, deletion, or modification of a variation point, or its variants takes place.
Changes can be recorded in change logs that can be created automatically by PAIS \cite{adaptiveprocess}. If a system does not provide a change log, as is typically the case of as non-process-aware information system, the change log needs to be extracted  from event logs by change mining \cite{handlingsudden}, \cite{detectingchange}. \\

\indent Change logs should provide information about the attributes of change in the process model or instance. The analysis of change logs can offer important information on changes that happened during the execution phase of a process, which can serve the re-engineering of the process models or can help learn from change and help identify the most appropriate change for future process instances \cite{usingprocessmining}, as well as evaluate the effect of the future changes before their application \cite{miningquerying}. The manual analysis of change logs is a hard task and change mining is meant to simplify the process of finding the required information by allowing rapid and effective detection of changes. In most cases changes are considered an unforeseen event in a business process event log, however a classification and recording of the reasons for and impact of the change are not available in existing solutions.

\subsection{Process Drift and Anomaly Detection}
\label{section:othermethods}
Process drift and anomaly detection are two areas related to process change mining since drifts and anomalies could also be considered as changes in business processes. Different aspects of dealing with concept drifts in processes are surveyed in \cite{sato2022survey}, providing an overview of the techniques and challenges. This study defines \textit{drift detection} as an approach that "detects that a process has changed without providing exact information about the time period or the trace/event the change occurred." \cite{KoC23} provides a systematic literature review of anomaly detection for business process event logs. It defines \textit{event log anomalies} as "anomalies in event logs are
process behavior that deviates from normal observations, whereby normal behavior can be defined in different ways depending on the application focus and the techniques
adopted to address the problem of anomaly detection."
However, in this study we consider only works whose target were changes with a purpose to improve process performance. Therefore process drift and anomaly detection are out of scope for this study. Naturally, in an additional line of research in future, these techniques could be considered in the overall process performance improvement cycle.

\section{Research Methodology}
\label{section:methodology}
\begin{figure}[h!]
\centering
\includegraphics[width=7cm]{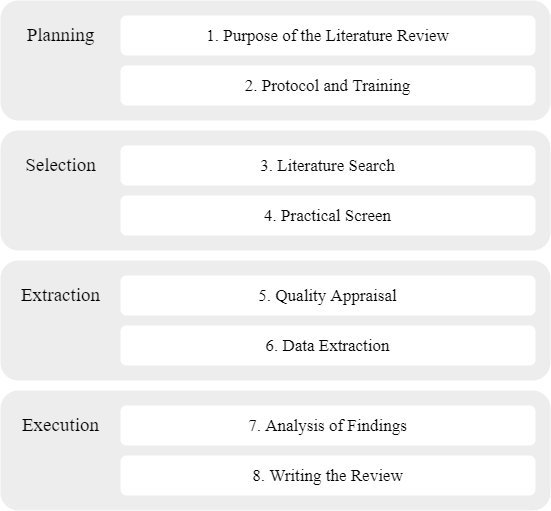}
\caption[Research Methodology]{Research Methodology (Following \cite{method})}
\label{fig:method}
\end{figure}

In this paper we carried out a systematic literature review based on the scientific guidelines specified in \cite{guide}. The guideline for the research methodology is outlined in \figurename~\ref{fig:method} and will be discussed in detail in the following sections. We selected this methodology based on its suitability for our objective, namely establishing the state of the art in change mining and logging in PAIS.

\indent  The research methodology requires to define a clear protocol for research after the purpose of the review has been identified. It is important that the criteria are defined clearly before the review is conducted, as a poorly conducted literature review can result in misconceptions. As a follow-up step, we conducted an extensive literature search based on the protocol.
Finally, we extracted and synthesized the data from the data set to write the review. \\
\indent The guideline in \figurename~\ref{fig:method} has eight elements that are required for a literature review as specified in \cite{guide2}. All of the steps are essential to the review and they will ensure that the review is scientifically rigorous \cite{guide2}.
In the following sections we present the individual steps of the methodology implementation.

\subsection{Research Questions}
\label{section:rq}
The aim of this paper is to review and evaluate the research in change mining and logging in process-aware information systems. To achieve this, we formulated research questions by using the PICOC (Population, Intervention, Comparison, Outcome, Context) criteria that are suggested by \cite{guide} as can be seen in Table \ref{table:PICOC}. In the criteria, population refers to the subjects that are affected by the intervention. Intervention refers to a tool that is directed at an issue. Comparison refers to another tool that is compared against the intervention. Outcome refers to factors that emerge as a result of the intervention. Context refers to conditions under which the intervention takes place.

\begin{table}[h!]
\centering
\begin{adjustbox}{width=\textwidth}
\begin{tabular}{ll}
\hline
\textbf{Criteria} & \textbf{Description} \\ \hline
Population & \begin{tabular}[c]{@{}l@{}}Literature on change mining and logging in process-aware information systems.\end{tabular} \\
Intervention & \begin{tabular}[c]{@{}l@{}} \\Extraction of methods and techniques that are used in change mining and \\logging in process-aware information systems.\end{tabular} \\ \\
Comparison &  N/A \\
Outcome & \begin{tabular}[c]{@{}l@{}} \\Describe the approaches used in change mining and logging in process-aware \\ information systems.\end{tabular} \\ \\
Context & Research to synthesize peer-reviewed literature.\\ \hline
\end{tabular}
\end{adjustbox}
\caption{PICOC Criteria for this review.}
\label{table:PICOC}
\end{table}

As a result, we formulated the following research question by following the criteria in Table \ref{table:PICOC}:
\begin{itemize}[label={RQ1.}]
  \item \textit{What primary studies have been published in the area of process-aware information systems that focus on change mining and logging?}
\end{itemize}

Since we want to focus on the content of the publications that we will find as a result of RQ1, we formulated four more research questions to help us to focus the research and investigate the technologies and methods that have been proposed in the publications.
\begin{itemize}[label={RQ2.}]
  \item \textit{How do we mine changes from business processes?}
\end{itemize}
\begin{itemize}[label={RQ3.}]
  \item \textit{How are changes that happened in business processes recorded?}
\end{itemize}
\begin{itemize}[label={RQ4.}]
   \item \textit{How are changes that happened in business processes stored?}
\end{itemize}
\begin{itemize}[label={RQ5.}]
    \item \textit{What do we learn from the recorded change information?}
\end{itemize}

RQ2 will help us to explore various change mining methods. RQ3 and RQ4 will provide us with information on how changes applied to processes are recognized, organized, and stored. RQ5 will helps us to discover different outputs of change mining approaches.

\subsection{Literature Research}
\begin{figure}[h!]
\centering
\includegraphics[width=12cm]{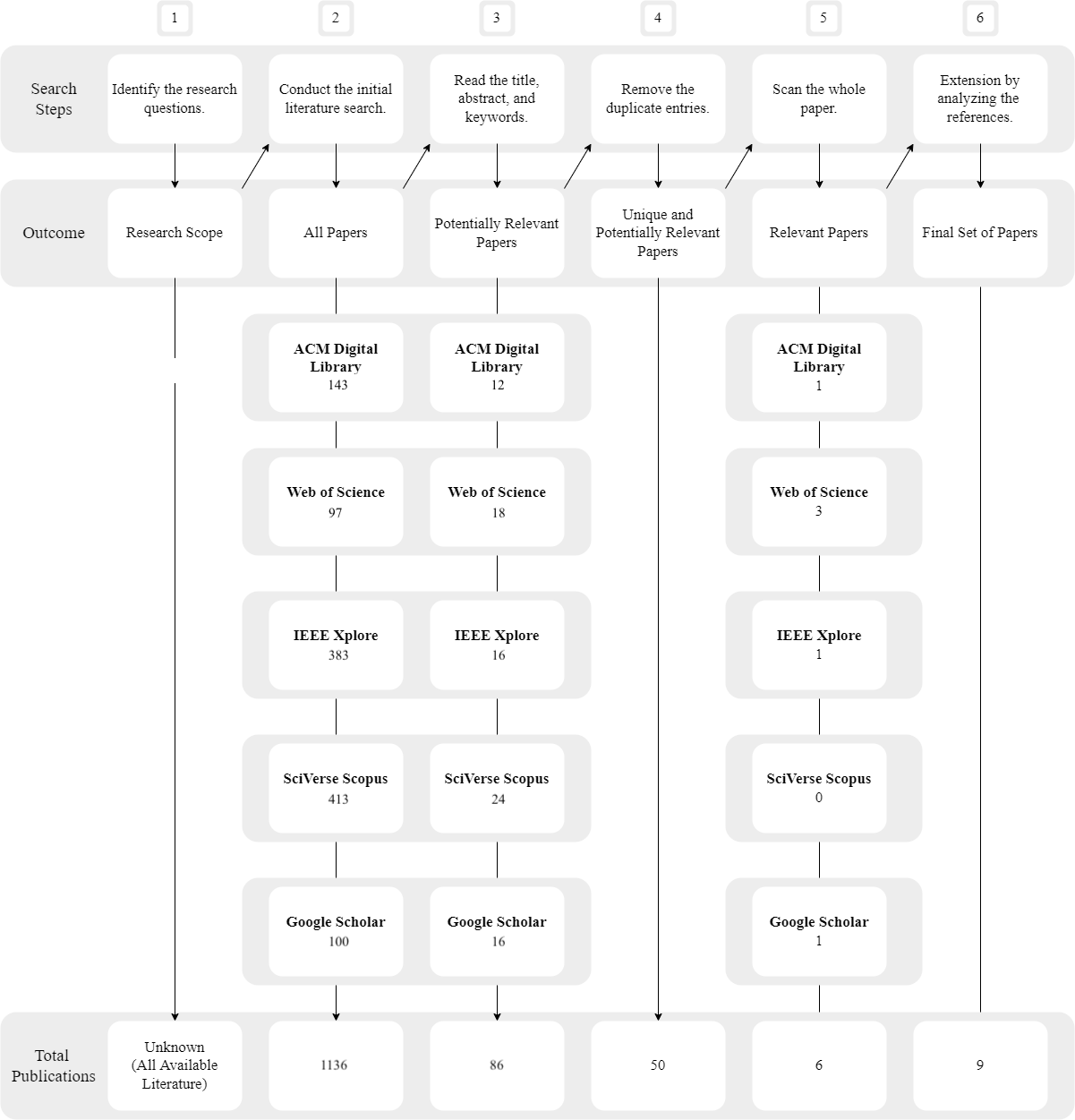}
\caption[Literature Search and Selection Process]{Literature Search and Selection Process (Following \cite{security,maturity})}
\label{fig:numbers}
\end{figure}

Our approach for the literature search was to find as many peer-reviewed publications as possible that can answer the research questions and narrow down the number of publications by using the selection criteria as can be seen in \figurename~\ref{fig:numbers}. To obtain the scientific publications for the review, we ran a manual horizontal search. The digital libraries of IEEE Xplore, Web of Science, SciVerse Scopus, ACM Digital Library, and Google Scholar were used to perform a horizontal search. The search strings that are used to carry the research are provided in Section \ref{section:string}. \\
\indent We started by searching for papers that are in the scope of our research. We narrowed down the initial set of papers by reading the title, abstract, and keywords on each of them. The general idea was to strictly keep the papers that used the keywords that are mentioned in Section \ref{section:string} in the title, abstract, or keywords. Then, we proceeded to remove the duplicate entries from the set of papers that we obtained. To remove the duplicate entries, we followed \cite{duplicate} which suggests that if a publication that has the same title and the same authors has been published in more than one digital library, they will be considered as one study and if the same authors published multiple publications on the same topic, the most recent one will be considered. Some of the papers were on specific areas of change in process-aware information systems so they had to be read in full to be categorized as relevant. Hence, we scanned the whole paper in the next step to further narrow down the set of papers and minimize the chance of removing a relevant publication. Publications that included any of the keywords but concentrated on areas that are not business process management and/or computer science were excluded. Furthermore, publications that focused on change propagation, concept drift, and process model discovery were excluded as they were not exactly the focus of our research, although the publications contained the search words. Finally, we were left with 6 publications that focused on change mining and logging in process-aware information systems, which we extended to 9 publications by examining their reference lists (see \figurename~\ref{fig:numbers}).

\indent After we determined the final set of papers, we designed a database to store the bibliographic information that we need from the publications. 

\subsubsection{Search Strings.}
\label{section:string}
%We start by defining a search string to conduct the literature review. 
To define the search string, we followed the guidelines suggested by Kitchenham et al. \cite{architectureguide} and as a result of this:
\begin{enumerate}[label=(\alph*)]
    \item “process-aware information systems”, “change”, “mining” and “logging” are the main search terms that were derived from the research question.
    \item “business processes”, “service composition”, “service orchestration”, “adaptation”, and “workflows” are technical synonyms that were derived from the main search terms.    
    \item The Boolean AND was used to connect some of the search terms in (a) and (b) to be able to narrow down the search results.
    \item The Boolean OR was used to incorporate some of the search terms in (a) and (b) to be able to broaden the search results.
    \item Some of the keywords in (a) and (b) were put in quotation marks to search for an exact match to be able to narrow down the search results.
\end{enumerate}

Finally, we combined all the terms and the guidelines that are mentioned above which led us to the following search string:
\begin{quote}
    \textit{((adaptation OR change) AND (logging OR mining) AND \\ (workflow OR “business process” OR “service composition” OR \\“service orchestration” OR “process-aware information systems”))}
\end{quote}

The keywords “information systems” and “processes” were considered in the initial string but they were removed after we observed that they brought irrelevant results. The keyword “ad-hoc change” was removed from the string as it did not change the result of the search.

\subsubsection{Search Sources.}
\label{section:source}
After we defined the search strings, the search sources had to be determined. It is important that we have a list of all the search sources that we used as this will ensure that other researchers, scholars, and practitioners will get the same search results from the search sources with the search strings that we use for the literature review. Table \ref{table:databases} shows the list of electronic databases that were used for the research of the review. The databases below were chosen as suggested by \cite{guide} \cite{architecture} because they cover most of the scientific publications in the field of computer science and guarantee to provide the confidence level for the inclusion of all the required primary studies. We applied the criteria defined in Table \ref{table:criteria} to all the digital libraries that are listed in Table \ref{table:databases}. 

\begin{table}[h!]
\centering
\begin{tabular}{@{}ccc@{}}
\toprule
\multicolumn{1}{c}{\textbf{Database}} & \multicolumn{1}{c}{\textbf{Institution}} & \multicolumn{1}{c}{\textbf{Abbreviation}} \\ \midrule
ACM Digital Library & ACM & ACM \\
Web of Science & Thomson Reuters & ISI \\
SciVerse Scopus & Elsevier & ELSV \\
IEEE Xplore & IEEE & IEEE \\
Google Scholar & N/A & N/A \\ \bottomrule
\end{tabular}
\caption{List of Electronic Databases}
\label{table:databases}
\end{table}

The ACM Digital Library provides the ability to extend the search by using the ACM Computing Literature Library which helped us to enrich the publications that we collected. Google Scholar brought up over 15000 results when we used the search string and narrowed down the search to publications that were published between the years 2000-2024. We decided to only cover the first 100 results that were listed according to their relevance to the search. SpringerLink was initially considered in the list of libraries to conduct the search as suggested by the guideline of Kitchenham et al. \cite{guide} but the results that we obtained were either irrelevant to the research or already found in the other digital libraries.

\subsection{Selection Criteria}
\label{section:criteria}
In order to ensure that the set of papers that we collected was relevant to the topic,  we determined a set of criteria which can be seen in Table \ref{table:criteria} to obtain the papers for the literature review. Publications that did not pass any of the criteria listed on Table \ref{table:criteria}were not considered for further evaluation and the publications were be excluded. Publications that paswere assessed further to decide if they propose a solution for the research questions. \\
\indent All of the digital libraries that are listed in Table \ref{table:databases} offered the option to exclude and include languages, publication types, publication years, and disciplines on the search which we utilized by using the criteria that are defined in Table \ref{table:criteria}. The use of the criteria helped us to find the key publications that would help us to answer the research questions and it ensured that the results are reliable and reproducible. \\
\indent Our criteria focus on content and publication. Content-related exclusion was applied in Steps 3 and 5 of the search process and publication-related exclusion was applied in Steps 2 and 6 of the search process which can be seen in \figurename~\ref{fig:numbers}.

\begin{table}[h!]
\centering
\begin{adjustbox}{width=\textwidth}
\begin{tabular}{@{}ll@{}}
\toprule
\textbf{Criteria} & \textbf{Description} \\ \midrule
\textbf{Inclusion Criteria} & Is the full-text of the study digitally accessible? \\
 & Is the full-text of the study in English? \\
 & Has the study been published between the years 2000 and 2024? \\
 & Does the title, abstract, and keywords of the study comply with the search strings? \\
 & Has the study been published in a scientific peer-reviewed source? \\ \\
\textbf{Exclusion Criteria} & Is the study in an area other than business process management and/or computer science? \\
 & Is the focus of the study not software systems, workflows, processes, or service compositions? \\ \bottomrule
\end{tabular}
\end{adjustbox}
\caption{Inclusion and Exclusion Criteria}
\label{table:criteria}
\end{table}

\subsection{Extending the Publication List}
Based on the literature search and selection process we identified 6 publications relevant for the research questions we defined in Section \ref{section:rq}. As a next step we looked into the references of the publications that we have already collected, as we already knew that they contained the information that was relevant to our research with the intention to extend the amount of available relevant literature. \\
\indent The procedure that we used to extend the set of publications with the references was similar to the original procedure and uses the same criteria that we have defined in Table \ref{table:criteria}.  
As a result we could add three more publications to the set of selected publications.

\subsection{Data Extraction and Synthesis}
\label{section:extract}
The publications that were collected had to be analyzed and structured to obtain the information that we needed to address the research questions. To collect the relevant information from the publications, we read the publications in detail and created lists of attributes related to each research question, and created a data extraction form. The list of identified attributes can be seen in Table \ref{tab:attributes}.

\begin{table}[h!]
\centering
\begin{adjustbox} {width=\textwidth}
\begin{tabular}{@{}ll@{}}
\toprule
\textbf{Research Question} & \textbf{Attributes} \\ \midrule
RQ1 & Title, Publication Year, Publication Type, Publication Source \\
RQ2 & Adapted Process Mining Methods, Novel Methods for Change Mining \\
RQ3 & Recording Individual Change Operations, Process Model Comparison \\
RQ4 & ADEPT Change Log, XML, MXML, CSV, XES \\
RQ5 & Change Model, Change Trees, Change Recommendations, Change Rule \\ \bottomrule
\end{tabular}
\end{adjustbox}
\caption{Attributes for Data Extraction}
\label{tab:attributes}
\end{table}

\indent In order to explore \textit{RQ1 - What primary studies have been published in the area of process-aware information systems that focus on change mining and logging?}, we collected a set of common fields such as the title, publication year, publication source, and publication type of the publications.\\
\indent In order to answer \textit{RQ2 - How do we mine changes from business processes?}, we chose the attributes that are listed below:
\begin{enumerate}
    \item \textit{Adapted Process Mining Methods:} In this class, we include the research works which adapt process mining methods to mine changes from business processes.
    \item \textit{Novel Methods for Change Mining:} In this class, we include research works that present novel methods for change mining. By novel, we mean the methods that are not extensions of process mining methods.
\end{enumerate}

In order to answer \textit{RQ3 - How are changes that happened in business processes recorded?}, we chose the attributes that are listed below: 
\begin{enumerate}
    \item \textit{Recording Individual Change Operations:} In this class, we include the research works that create change logs by recording each change operation.
    \item \textit{Creating a Change Log Using Process Model Comparison:} In this class, we include research works that compare two versions of a process model and extract the changes that the model has gone through in order to create a change log.
\end{enumerate}

In order to answer \textit{RQ4 - How are changes that happened in business processes stored?}, we chose the attributes that are listed below:
\begin{enumerate}
    \item \textit{ADEPT Change Log:} In this class, we include research works that use ADEPT change logs for change log storage.
    \item \textit{XML:} In this class, we include research works that use XML for change log storage.
    \item \textit{MXML:} In this class, we include research works that use MXML for change log storage. This includes works that convert other formats into MXML as a preparation step for different procedures as well.
    \item \textit{CSV:} In this class, we include research works that use CSV for change log storage.
    \item \textit{XES:} In this class, we include research works that use XES~\cite{xesstandard} for change log storage, which is the standardized format for the representation of event logs.
\end{enumerate}

In order to answer \textit{RQ5 - What do we learn from the recorded change information?}, we chose the attributes that are listed below:
\begin{enumerate}
    \item \textit{Change Model:} In this class, we include research that creates a model of the mined changes. These models can be represented in different formats such as Petri nets or BPMN diagrams and show the dependencies between change operations.
    \item \textit{Change Trees:} In this class, we include research works that create a change tree from the mined changes in order to visualize the frequencies and dependencies of different change operations.
    \item \textit{Change Recommendations:} In this class, we include research works that focus on generating a list of recommended changes to be applied to processes.
    \item \textit{Change Rule:} In this class, we include research works that focus on the creation of change rules with the use of process change information and contextual information. 
\end{enumerate}

Table \ref{tab:form} depicts the form available online\footnote{The form with the information that is extracted from the publications can be accessed at: \url{https://docs.google.com/spreadsheets/d/1ZFXsqRkkknaQHfENsWkoe0iECXFXJfd4LPCmMRmtLwg/}} that we have used to extract data from our final set of publications. The form contains the information for each publication. 

\begin{table}[h!]
\centering
\begin{adjustbox} {width=\textwidth}
\begin{tabular}{@{}ll@{}}
\toprule
\textbf{Data Item} & \textbf{Aim} \\ \midrule
Title and Reference & Provide full title and publication details of the paper. \\
Aim of Research & Summarize what researchers aimed to achieve in the paper. \\
Relevant Research Questions & Identify which of the research questions are answered in the paper. \\
Relevant Attributes & Identify which of the attributes in Table \ref{tab:attributes} exist in the paper. \\
Research Contribution & Summarize contributions of the paper to the field. \\
Implementation Details (If Applicable) & Explain the use of tools, datasets, and techniques. \\
Conclusion & Summarize the conclusions that are stated by the authors. \\ \bottomrule
\end{tabular}
\end{adjustbox}
\caption{Literature Review Form}
\label{tab:form}
\end{table}

\section{Results}
\label{section:results}
In this section we  will present our literature review finding. First, we will outline the bibliographic information about the publications from the final set. Then, we will present  the findings and the answers the research questions.

\subsection{Overview of Selected Publications}
\label{section:overview}
In this section, we provide an overview of the publications by presenting the publication types, publication sources, and publication years. \\
The list of selected publications is shown in \ref{tab:names}.

\begin{table}[h!]
\centering
\begin{adjustbox}{width=\textwidth}
\begin{tabular}{@{}ccl@{}}
\toprule
\textbf{Reference} & \textbf{ Year } & \textbf{Title} \\ \midrule
\cite{applyingchange} & 2022 & Applying Change Tree Method in Configurable Process Mining \\
\cite{handlingsudden} & 2021 & Handling Sudden and Recurrent Changes in Business Process Variability: Change Mining Based Approach \\
\cite{enhancing} & 2021 & Enhancing Change Mining from a Collection of Event Logs: Merging and Filtering Approaches \\
\cite{anewframework} & 2020 & A New Framework to Improve Change Mining in Configurable Process \\
\cite{waas} & 2015 & Mining Change Operations for Workflow Platform as a Service \\
\cite{detectingchange} & 2015 & Detecting Change in Processes Using Comparative Trace Clustering \\
\cite{miningquerying} & 2015 & Mining and Querying Process Change Information Based on Change Trees \\
\cite{usingprocessmining} & 2008 & Using Process Mining to Learn from Process Changes in Evolutionary Systems \\
\cite{purging} & 2006 & On Representing, Purging, and Utilizing Change Logs in Process Management Systems \\ \bottomrule
\end{tabular}
\end{adjustbox}
\caption{List of Primary Studies}
\label{tab:names}
\end{table}

\indent Table \ref{tab:yeartype} gives an overview of the publication types of the selected publications and their publication years, and shows that the final set consists of 5 conference papers and 4 journal articles published between the years 2000 and 2024. 

\begin{table}[h!]
\centering
\begin{tabular}{@{}llllllll@{}}
\toprule
 & \textbf{2022} & \textbf{2021} & \textbf{2020} & \textbf{2015} & \textbf{2008} & \textbf{2006} & \textbf{Total} \\ \midrule
\textbf{Conference} & 1 & 0 & 1 & 2 & 0 & 1 & 5 \\
\textbf{Journal} & 0 & 2 & 0 & 1 & 1 & 0 & 4 \\
\textbf{Total} & 1 & 2 & 1 & 3 & 1 & 1 & 9 \\ \bottomrule
\end{tabular}
\caption{Summary of Publication Years}
\label{tab:yeartype}
\end{table}

Table \ref{tab:sourceyear} lists the publication sources of the selected publications and their publication years.

\begin{table}[h!]
\centering
\begin{adjustbox}{width=\textwidth}
\begin{tabular}{@{}lll@{}}
\toprule
 & \textbf{Journals} & \textbf{Conferences} \\ \midrule
\textbf{2022} &  & \begin{tabular}[c]{@{}l@{}}IRASET'2022: 2nd International \\ Conference on Innovative Research in \\ Applied Science, Engineering and Technology\end{tabular} \\ \\
\textbf{2021} & \begin{tabular}[c]{@{}l@{}}International Journal of Advanced\\ Computer Science and Applications\end{tabular} & \\ \\
\textbf{} & Journal of Physics Conference Series &  \\ \\
\textbf{2020} &  & \begin{tabular}[c]{@{}l@{}}NISS2020: Proceedings of the 3rd International\\ Conference on Networking, Information\\ Systems \& Security\end{tabular} \\ \\
\textbf{2015} & World Wide Web & \begin{tabular}[c]{@{}l@{}} ICSOC'2015: International Conference on \\ Service-Oriented Computing \end{tabular} \\ \\
\textbf{} &  & \begin{tabular}[c]{@{}l@{}}SIMPDA'2015: 5th International Symposium on\\ Data-Driven Process Discovery and Analysis\end{tabular} \\ \\
\textbf{2008} & \begin{tabular}[c]{@{}l@{}}International Journal of Business\\ Process Integration and Management\end{tabular} &  \\ \\
\textbf{2006} &  & \begin{tabular}[c]{@{}l@{}}BPM'06: Proceedings of the 4th International\\ Conference on Business Process Management\end{tabular} \\ \bottomrule
\end{tabular}
\end{adjustbox}
\caption{Summary of Publication Sources}
\label{tab:sourceyear}
\end{table}

\subsection{Overview of Primary Studies' Contributions}
\label{section:findings}
In this section, we present the summaries of the contributions of the selected publications with the help of the data extraction form. For more details of the selected works, the reader is referred to the publications themselves and to comprehensive summaries of the works presented in \cite{hande}.

\textit{Several publications of Hmami et al. \cite{applyingchange,handlingsudden,enhancing,anewframework} }propose a framework for change mining in business processes, which they claim can be used as a foundation for a recommendation system to help the business process designers change processes in order to satisfy new requirements. In \cite{anewframework} the authors introduce the overall framework with focus on change mining from a collection of event logs in order to recognize and recommend changes in business processes. The authors state that the use of a collection of event logs for change discovery of a configurable process that is related to a family of process variants is difficult due to the fact that such collections require the identification of the variability specification of the collection and discovery of changes of the variable elements when a mining approach is applied to them. The framework that they propose consists of four components:
\begin{enumerate}
    \item \textit{The First Component} takes a set of similar event logs as input and merges them into one event log, presented in \cite{enhancing}. The output contains common parts and variable parts in the same log.
    \item \textit{The Second Component} reduces the volume of the output data from the first component by filtering the variable parts from the common parts, which results in a variability event log. Hmami et al. \cite{enhancing} provide a filtering approach which relies on the so-called variability specification file. The variability specification file indicates the variable fragments which are to be mined in order to discover variability changes. In \cite{enhancing}, an event log of variable fragments is created from a collection of event logs and is to be used as an input for change mining. 
    \item \textit{The Third Component} compares the variability event log and variability specification file to highlight the change points and outputs a variability change log that contains the changes that were applied to the variable elements. This component of the framework is realized in detail by the approach presented by the same authors in \cite{handlingsudden}. There, a change mining approach is introduced,  inspired by a machine learning technique called STAGGER, and extracts the changed fragments from a set of events. The extracted change fragments are stored in an XML-based change log.
    \item \textit{The Fourth Component} generates a list of changes from the variability change log, which are recommended for the business process models in the future, to business process designers. A so called change tree method \cite{applyingchange} is used by the authors to represent events in a change log as a tree.
\end{enumerate}

\indent In this work, the authors assume that an event log can be represented as a CSV file, MXML (Mining Extensible Markup Language), or XES (Extensible Event Stream). CSV was used by Hmami et al. in \cite{enhancing} to make the implementation of the proposed method more straightforward. \\
\indent Hmami et al. emphasize the importance of detection of changes before a process mining approach is applied to an event log. Thus, in \cite{handlingsudden}, the focus is on the detection of sudden and recurrent changes. According to the same work \cite{handlingsudden} a sudden change is an unforeseen event that occurs unexpectedly, whereas a recurrent change is a seasonal change that can recur multiple times.

In order to test the proposed approach, researchers have implemented the method on the toolset “Random Configurable Process Model Generator'', where they have previously implemented the merging and filtering steps of the proposed framework. The results have shown that the proposed approach by Hmami et al. \cite{handlingsudden} has the ability to detect drift on synthetic event logs. However, tests on a real collection of event logs have not been carried out yet. 

\textit{The work of Cao et al.~\cite{waas}} proposes an approach for  mining workflow changes of users in WaaS (Workflow Platform as a Service) environment with the goal to discover change rules and reuse these rules to adapt workflow models to different situations automatically in the future. The proposed approach has three phases:
\begin{enumerate}
    \item A workflow model will be specified and deployed to the WaaS.
    \item Changes applied to the original workflow model will be retrieved and used for the discovery of change rules.
    \item Discovered change rules will be automatically applied to the workflow model based on the context information.
\end{enumerate}

In their work, Cao et al. \cite{waas} state that a process model can be converted to a well-structured model \cite{graphrefactoring}, and a well-structured process model can be converted into a tree, which is known as a Process Structure Tree (PST) \cite{refinedprocess}. Based on that, \cite{waas} compares PSTs of two versions of a process model in order to retrieve change operations applied to the original version. The approach is supposed to be used instead of the registration of all change operations which is hard to analyze and reuse if there are too many operations. The authors claim that it is simpler to reuse the retrieved changes as they are organized as a comparison to the records of all changes. Detection of differences between two PSTs requires the largest common subtree to be found. This problem is called the Maximum Common Subtree Isomorphism problem \cite{algorithmsontrees}. However, available Maximal Common Subtree (MCST) isomorphism methods cannot be used for PSTs, as PSTs are semi-ordered trees. In \cite{waas}, the authors revise the algorithm from \cite{editdistance} which computes constraint edit distances between two semi-ordered trees in order to solve the top-down MCST isomorphism problem in PSTs. Based on MCSTs, they provide a change operation retrieval algorithm, which assumes that all deletions occur before all insertions in order to avoid the retrieval of redundant operations. The retrieval algorithm takes the original PST and the revised PST as input. 

The authors \cite{waas} state that the retrieved change information can be utilized to mine rules for process model configuration. For example, a decision tree can be used to generate reuse rules for individual change operations by transformation of a path into a rule, for which they define a corresponding algorithm. The proposed approach has been implemented as a prototype.

\textit{In publication \cite{detectingchange} by Hompes et al}., a comparative trace clustering technique is introduced to detect changes in the behavior of a business process by comparing changes in clusters over sub-logs. This work \cite{detectingchange} extends the work in \cite{discoveringdeviating} to present a new method for the detection of change points. The method presented in \cite{discoveringdeviating} combines outlier detection and trace clustering for the detection of deviations and common behavior, and uses the MCL (Markov Cluster) algorithm \cite{clusteralgorithm} for that. The method in \cite{detectingchange} introduces the detection of changes in control-flow and data between the different executions of a business process and is implemented in ProM, available as a package called TraceClustering\footnote{https://promtools.org}.\\
\indent In the method proposed in \cite{detectingchange}, a stochastic similarity matrix between the different executions of a business process is used as an input in MCL. Expansion and inflation operations are alternated repeatedly on the similarity matrix in order to achieve the separation of the matrix into distinct elements, which results in clusters. In expansion, the matrix is raised to a certain power and in inflation, every element is raised to a certain power and the matrix is normalized so that it remains stochastic. Similarities between different executions of business processes are computed by mapping each execution to a profile vector with the use of perspectives and calculation of the values of pair-wise vector similarities. Perspectives can be the frequency of specific activities, execution data, event data, or such. As new events happen in a specific trace, the similarity of that trace to other traces will be changed, which can be captured in the similarity matrix. The change of values in the similarity matrix is computed over time to find important change points. For each event window defined over time, executions of business processes that contain events either in that window or before that window are used in the computation of the next similarity matrix. Attributes of process executions are assumed to be known from the beginning of a trace of an execution. Therefore, change points will be visible at the beginning of an execution that deviates and will result in a sub-log of all events that occurred before a specific time. For that sub-log, a similarity matrix is constructed and compared with the similarity matrix from before. If the change points are known, the behavior before and after the change points can be compared by using the clusters of process executions based on the time of occurrence relative to the change points, which can reveal the characteristics of the behavior that has changed. There will be a separate cluster for a behavior that became less common and that will not be put in the same cluster as another common behavior.

\textit{In publication \cite{miningquerying} by Kaes and Rinderle-Ma}, two representation methods for change log data -- change trees and n-gram change trees -- are presented and used  to analyze changes in adaptive process instances. The authors state that analysis of similarities between evolutions of process instances can serve as a basis for the prediction of future changes. To visualize which process instances evolved similarly, a representation should contain changes made to all instances chronologically. In change trees, the order of change operations from change instances is preserved and represented along paths that go from the root of the tree to the leaves. The methods are implemented as a plugin in ProM and uses MXML and XES log files. 
The main advantage of change trees compared to other change mining tools is that they visualize various change instances and change occurrences that are observed in highly adaptive process scenarios. In addition to that, n-gram change trees provide information on changes that occurred after a specific change sequence. 
In the same work, Kaes and Rinderle-Ma \cite{miningquerying} work with the assumption that change instances can be treated as strings and change sequences can be treated as substrings, which transforms the problem at hand into the problem of n-gram models in language processing. Thus, in such a context, the n-gram will represent the change pattern of interest and enable the identification of the change sequences that follow the change pattern for all occurrences in the change log. In n-gram change trees, a change gets appended to the n-gram change tree if it occurs after the application of the n-gram to the instance compared to regular change trees.

\textit{The work of Günther et al. \cite{usingprocessmining}}, recognizes three challenges  for providing a framework for mining ad-hoc changes in adaptive process management systems, which are:
\begin{enumerate}
    \item Determination of runtime information that needs to be logged and the optimal representation of the selected information in order to achieve better mining results.
    \item Development of advanced mining techniques which use change logs together with execution logs.
    \item Integration of the developed mining techniques with the already existing adaptive process management technology. 
\end{enumerate}

Günther et al. propose two new methods in \cite{usingprocessmining} for ad-hoc change mining in adaptive process management systems. The proposed methods are implemented as plug-ins in ProM and discover change knowledge from ADEPT change logs mapped to MXML \cite{promframework,ageneric} format. They acquire an abstract change process as a Petri net model, which shows the changes made on instances of a specific process type. In order to map ADEPT change logs into MXML, Günther et al. propose the use of ProMimport, which converts data from various systems to MXML \cite{ageneric}.The authors also state that change logs can be converted to and represented by MXML with slight modifications. For instance, the attributes that define a change event are not considered in the standard MXML format but rather treated as as attributes in the “Data” field in the MXML change log. These attributes are \cite{usingprocessmining}:
\begin{enumerate}
    \item The type of change.
    \item The subject that mainly got affected.
    \item The syntactic context of change which can be defined as the elements in the process model that directly precede or follow the subject of the change.
\end{enumerate}

The authors also suggest for the attributes of a change event to  include a rationale field as well, which would describe the incentive behind the change. The field that indicates the person, who has applied the respective change, would be the originator field. The exact date and time of the change is indicated by the timestamp field. Change events can be seen as atomic operations and for that reason, the event type is “complete” by default. Process elements must be indicated for each change event in order to achieve backward compatibility of MXML change logs with the conventional process model algorithms. However, such information is unavailable as the change process does not adhere to an a-priori process model. Therefore, a combination of the change type and subject is used in \cite{usingprocessmining} instead in order to uniquely identify classes of changes. Conversion of change log data to MXML allows the usage of conventional process mining algorithms on change logs, which are evaluated by the authors. \\
\indent For the evaluation, Günter et al. used an extension of the ADEPT demonstrator and compared algorithms based on quality criteria. They have analyzed the $\alpha$-Algorithm, the MultiPhase Miner, and the Heuristics Miner and found that these algorithms capture dependencies between change operations successfully for simple processes and a small set of change operations. However, the quality of the change processes decreases when different change operations are used and the complexity of processes increase. In such a scenario, non-existent dependencies are generated by the mining algorithms and change processes become less accurate. The main source of this problem is the fact that change logs are not as populated as enactment logs. Therefore, they contain little data to learn from. \\
\indent The first method \cite{usingprocessmining} proposes is based on MultiPhase mining \cite{multiphase} and uses additional information related to the semantics of change operations. Günther et al. \cite{usingprocessmining} selected the MultiPhase algorithm as a basis for their contribution, because change process mining is similar to process mining from enactment logs. According to the authors, the MultiPhase algorithm stood out from the other available process mining algorithms as it handles complex branchings robustly, but they note that any other process mining approach which detects causalities explicitly could be used. Based on the casualties derived from the log, the MultiPhase mining algorithm can construct basic workflow graphs, EPC models, and Petri nets. The MultiPhase mining algorithm generates a model for each process instance which only needs to capture causal dependencies as there are no choices in a single instance \cite{usingprocessmining}. Then, the generated models are aggregated to create a general model for the set of change logs. As van Dongen and van der Aalst \cite{multiphase} state, the causal relations are inferred from the change log as follows in the MultiPhase algorithm: "\textit{If a change operation A is followed by another change B in at least one process instance, and no instance contains B followed by A, the algorithm assumes a possible causal relation from A to B.}" \\
\indent Günther et al. introduce the concept of commutativity-induced concurrency to extend MultiPhase mining by removing unnecessary causal relations that do not reflect dependencies between change operations. This in return ensures that not every two change operations are required to be observed in both orders to be found concurrent, which is specifically important in change logs since they usually contain fewer data compared to execution logs. \\
\indent The second method Günther et al. propose is based on the theory of regions \cite{derivingpetrinets} and generates a Petri net model by mapping change logs to a labeled state transition system. This method builds upon the fact that a process model can be reconstructed from the original model and a sequence of change operations. A transition system is constructed based on the fact that a sequence of changes defines a state and the assumption that the changes are memoryless, which means that the process model after the change contains the necessary data. Every process model visited in the change log is represented by the states in the transition system. The theory of regions provides the functionality to map the transition system to a change process model as a Petri net. In such a context, Petrify \cite{petrify} can be utilized to construct Petri nets for transition systems with the use of regions. \\
\indent Günther et al. state that the proposed methods provide an overview of instance changes that have been made at the system level, to the process engineers. They conclude that the methods produce different results as they are different ways of evaluating change processes but add that the first method performs better on small quantities of change log instances. The second method performs better when large quantities of change log instances are available as it has better accuracy in capturing the observed sequences of changes.

\textit{In publication \cite{purging} by Rinderle et al.}, the authors present the set of information related to changes that should be stored in change logs, and a method for the efficient representation of the stored changes, which is implemented in ADEPT2\footnote{ADEPT2 is an adaptive, high-performance process management tool.}.
The motivation for the work \cite{purging} is that the type of logged change information and the representation of logged change information affect the usefulness of the change logs. Change logs often contain noise, which is unnecessary, irrelevant, or is simply wrong information and has impact on change mining, change effect analysis, and comparison of changes, and change log management should allow different views of the log information for such use cases. For instance, only the necessary information should be provided for change mining and conflict detection, as noisy or irrelevant information makes it difficult to check or compare changes in processes. However, the presentation of all change transactions, whether necessary or irrelevant, is required for traceability purposes. Hence, the authors identify four use cases for change log management and discuss in which situations purged change logs are more useful than noisy change logs along these use cases:
\begin{enumerate}
    \item Restoration of the logical structure of a process instance that has undergone ad-hoc changes by providing additional information to be used with event log information.
    \item Enablement of change traceability, which is a critical requirement as all changes applied to the standard procedures have to be documented for legal reasons in specific domains.
    \item Reuse of change if similar situations happen again.
    \item Detection of conflicts between concurrent changes applied to a process from change log information.
\end{enumerate}

\indent In order to meet the requirements of the aforementioned use cases, a suitable representation for the change log information and methods to process it must be determined. Rinderle et al. recognize two basic approaches for defining the changes in process graphs which are:
\begin{enumerate}
    \item \textit{Application of Graph Primitives:} Changes are defined on process graphs by application of sequences of graph primitives, such as insertion or deletion of nodes and edges. This approach would allow restoration of process structures, and enable effective conflict checks, but results in loss of information related to the semantics of the changes, which limits change traceability and conflict analyses.
    \item \textit{By High-Level Change Operations:} Changes are defined on process graphs with the use of high-level operations which combine change primitives in specific ways. This approach contains more change semantics and is associated with formal preconditions and postconditions. High-level change operations can be grouped together in change transactions in order to express complex changes, which may be needed if the application of a set of concomitant change operations is required.
\end{enumerate}

Furthermore, the same work \cite{purging} proposes an algorithm for purging change logs from the noise, which can be applied to complex change transactions as well as simple ones. 
The concept of the delta layer is also introduced in this publication  to record the deviations of process instances from the process template  and contains only the changed parts of the process template in a specific instance. Instance-specific change information is then obtained by using a process template object together with the delta layer object thus maintaining the traceability of changes at the type level and instance level.

\section{Discussion of Findings}
\label{section:discussion}
This section will present an analysis of metadata of the publications, elaborate on the results and their relevance to our research questions, and discuss the limitations of the review.

\subsection{Discussion of Publications}
In our final set of publications, 5 of the papers were published in 3 recent consecutive years, with 3 of them being conference papers, which might indicate a trend of increasing interest in change mining by the community. Nevertheless, we also observe that none of our selected papers was published in the same journal or the same conference as any other one. We cannot help noticing that the publications in the most recent years have not been published at any of the outlets common for the BPM/workflow management community.

\subsection{Discussion of Findings}
In this section we will present our findings on methods for change logging and mining change logs in process-aware information systems, based on the 9 publications, published between the years 2000 to 2024 (see Table \ref{tab:names}) that we selected following the SLR methodology.  Recall that these are the studies that represent the answer to \textit{RQ1}: \textit{What primary studies have been published in the area of process-aware information systems that focus on change mining and logging?}.

We analyzed and classified the publications to answer the rest of the research questions listed in Section \ref{section:rq}. The findings of the review in regards to the other research questions are as follows:
\begin{enumerate}

    \item [\textit{RQ2:}] \textit{How do we mine changes from business processes?} 
    
    We have recognized two main categories of methods for change mining:
    \begin{enumerate}
        \item \textit{Adapted Process Mining Methods:} In this class, we have included the methods that utilize the application of process mining methods for change mining purposes. Application of $\alpha$-Algorithm, MultiPhase Miner, and Heuristics Miner algorithms to MXML change logs has been presented in this category. However, so far only MultiPhase Miner has been optimized amongst them to handle complex processes and various changes \cite{usingprocessmining}. Application of trace clustering based methods for change mining purposes \cite{detectingchange} are considered in this category as well.
        \item \textit{Novel Methods for Change Mining:} In this class, we have included the methods for change mining that are not extensions  of process mining methods. The methods in this class are mapping of change logs to labeled state transition systems \cite{usingprocessmining}, algorithms based on machine learning \cite{handlingsudden}, and methods based on comparing process structure trees \cite{waas}.
    \end{enumerate} \vspace{0.5em}
    
    \item [\textit{RQ3:}]  \textit{How are changes that happened in business processes recorded?} 
    
    We have recognized that changes in business processes are recorded into change logs either by recording each change operation \cite{waas} or identified by comparing original and revised process models \cite{waas}. It is noteworthy that sometimes change information does not get recorded at all and has to be extracted from event logs \cite{anewframework}. The problem of noisy change logs is also relevant here and some approaches remove such logs when all change operations are recorded \cite{purging}. We would like to emphasize that despite using the standard SLR methodology, the articles selected contain little information on how changes are recorded. Recording of changes is a task that must be handled by process-aware information systems, and only \cite{waas} and \cite{purging} were written from the perspective of a process-aware information system.  \vspace{0.5em}
    
    \item [\textit{RQ4:}] \textit{How are changes that happened in business processes stored?}
    
    We have recognized five formats for storage of change logs which are ADEPT change logs \cite{usingprocessmining}, XML \cite{handlingsudden,applyingchange}, MXML \cite{usingprocessmining,miningquerying}, CSV \cite{enhancing}, and XES \cite{miningquerying}. The works \cite{waas,anewframework,detectingchange,purging} do not specify a format for storage of change logs.  \vspace{0.5em}
    
    \item [\textit{RQ5:}] \textit{What do we learn from the recorded change information?}
    
    We have recognized four ways to learn from changes in business processes:
    \begin{enumerate}
        \item \textit{Change Model:} Change models are created from change logs in order to visualize the changes that have been applied to instances of a process \cite{usingprocessmining}.
        \item \textit{Change Trees:} Change trees are used for the presentation of the information in the change logs \cite{applyingchange,miningquerying}. The advantage of change trees is that they provide a view of various change instances along with their occurrences, and reveal the similarities between changes applied to different process instances.
        \item \textit{Change Recommendations:} A list of change recommendations to be applied to a process can be generated from change information in the context of business process variability \cite{anewframework}, which enables the enhancement of business processes.
        \item \textit{Change Rule:} Change rules can be generated from change information and can be used for auto-configuration of processes based on context information \cite{waas}.
    \end{enumerate}
\end{enumerate}

\subsection{Limitations}
We have considered only peer-reviewed sources in this review. Naturally, there is a possibility that valuable information about change mining is available in sources such as websites of organizations that work on business process management, and documentations of change mining tools. If this is indeed the case, the review can be extended to retrieve such information in the future by repeating the review methodology described in this paper. \\
Based on the exclusion criteria we also did not consider publications with focus on concept drift in business processes and anomaly detection as these works have a different focus, however (parts of) some of them  might be of relevance to some of our research questions.

\section{Summary and Conclusions}
\label{section:conclusion}
This literature review aimed to analyze the methods that are used to store,
record and mine changes in business processes and investigated what current practices learn from the results of the change mining. To explore the field, we used a standard  review methodology that fit our objectives and was suitable towards answering the identified research questions. After the initial search, we examined 1136 peer-reviewed publications that were published between the
years 2000 and 2024. Based on the criteria we selected for the filtering of relevant publications our final selection included 6 papers. We extended the publication list by examining the references of the  6 papers that we initially selected. We ended up with 9 publications that had to be analyzed and synthesized to answer our research questions. We extracted the data of each publication by using a data extraction form (Table \ref{tab:form}) and presented the results of the research.\\

\indent We also recognize the higher maturity level if process mining techniques than the ones specifically targeting change mining. Therefore, the identification of commonalities and differences between change mining and process mining may enable the utilization of more process mining techniques in change mining. For future work, we see the need to study and compare other available process mining methods and potentially extend them for change mining. There is also a need of a comparative study to  assist practitioners to decide on methods to use for the development of PAIS as well. In addition, we believe that there may be machine learning algorithms, other than STAGGER \cite{handlingsudden}, that can be utilized for change mining purposes. Identification of methods similar to STAGGER and evaluation of whether they can be used in change mining or not might lead to the discovery of more efficient methods. Furthermore, we conclude that there is still room for improvement in change logging and mining that would lead to learning from adaptation and the impact of adaptation on process performance as measured by different KPIs. In particular, it is important to be able to measure the impact of changes in process instances as a result of an adaptation, record this fact and use it for change mining and subsequent recommendation for adaptations in future process instances.

\end{document}